# Stability Dust-Ion-Acoustic Wave In Dusty Plasmas With Stream -Influence Of Charge Fluctuation Of Dust Grains

Barbara Atamaniuk and Krzysztof Żuchowski

*Institute of Fundamental Technological Research, Polish Academy of Sciences,*
*00-049 Warsaw, Świętokrzyska 21,POLAND*

**Abstract.** There is a quickly increasing wealth of experimental data on so-called dusty plasmas i. e. ionized gases or usual plasmas that contain micron sized charged particles. Interest in these structures is driven both by their importance in many astrophysical as well as commercial situations. Among them are linear and nonlinear wave phenomena. We consider the influence of dust charge fluctuations on stability of the ion-acoustic waves when the stream of particles is present. It is assumed that all grains of dust have equal masses but charges are not constant in time-they may fluctuate in time. The dust charges are not really independent of the variations of the plasma potentials. All modes will influence the charging mechanism, and feedback will lead to several new interesting and unexpected phenomena. The charging of the grains depends on local plasma characteristics. If the waves disturb these characteristic, then charging of the grains is affected and the grain charge is modified, with a resulting feedback on the wave mode. In case considering here, when temperature of electrons is much greater then the temperature of the ions and temperature of electrons is not great enough for further ionization of the ions, we show that stability of the acoustic wave depends only one phenomenological coefficient and relation betwee.

## INTRODUCTION

In ideal dusty plasmas, where all grains of dust have the same atomic numbers $Z_d$ and masses $m_d$, exist low frequency modes DIAW (Dust Ion Acoustic Wave) and DAW (Dust Acoustic Wave) described by fluid [1], [2] or kinetic model [3]. On the other hand in a real Dusty Plasmas charges of grains can fluctuate absorbing or emitting some electrons[4], [5]. We assumed that fluctuation $\delta Z_d$ is the same for all grains but: $\delta Z_d \ll Z_d$, and a masses of grains $m_d$ are a constant. As a result of fluctuating dust charges in dusty plasmas, many new problems can appear which are in general treatment by Verheest [4]. We consider a specific problem when the temperature of electrons is much greater than the temperature of ions and temperature of electrons is not great enough for further ionization of the ions. Because the magnetic field is absent we conclude that our problem is onedimensional. In this case the continuity equations for specimens of dusty plasmas (grains of dust, ions, and electrons) can be written in the form:

$$\frac{\partial n_d}{\partial t} + \frac{\partial (n_d u_d)}{\partial x} = 0, \qquad (1)$$

$$\frac{\partial n_i}{\partial t} + \frac{\partial (n_i u_i)}{\partial x} = 0, \qquad (2)$$

$$\frac{\partial n_e}{\partial t} + \frac{\partial (n_e u_e)}{\partial x} = S_e. \qquad (3)$$

Here $n_d, n_i, n_e, u_d, u_i$ and $u_e$ -are grains of dust number density, ions number density, electrons number density, grains of dust fluid velocity, ions fluid velocity and electron fluid velocity. We assume, in a view above remark, that at the equilibrium level fluctuations of ions are absent. Due to the possible fluctuations of the dust charges we can express the conservation of charge in the dusty plasma by:

$$\frac{\partial}{\partial t}(-n_e e + n_d q_d + n_i e) + \frac{\partial}{\partial x}(-n_e e u_e + n_d q_d u_d + n_i e u_i) = 0, \qquad (4)$$

where: $q_d$ - the charge of grain of dust, $e$ - is the elementary positive charge. This can be rewritten with the help of the continuity equations (1)-(3) as:

$$n_d \left( \frac{\partial}{\partial x} + u_d \frac{\partial}{\partial x} \right) q_d = e S_e. \qquad (5)$$

On the other hand, the charge of grain of dust fluctuation is given by:

$$\frac{dq_d}{dt} \left( \frac{\partial}{\partial t} + u_d \frac{\partial}{\partial x} \right) q_d = I_i(n_i, q_d) + I_e(n_e, q_d), \qquad (6)$$

where $I_i(n_i, q_d)$ and $I_e(n_e, q_d)$ are the ionic and electronic charging current, respectively. When we combine (5) and (6), we get

$$e S_e = n_d I_e(n_e, q_d) + n_d I_i(n_i, q_d). \qquad (7)$$

In equilibrium dusty plasma, the total charging current vanishes:

$$I_{i0} + I_{e0} = 0, \qquad (8)$$

where $I_{i0}$ and $I_{e0}$ denotes the equilibrium charging current for ions and electrons, respectively. Therefore we can expand (7) as a function of $n_e, q_d$ and $n_d$ using (8) and hence in linear approximation for $S_e$ vanishing at equilibrium, it is given by:

$$S_e = -\nu_e \delta n_e - \mu_e \delta q_d, \qquad (9)$$

where $\nu_e, \mu_e$ denotes charging fluctuation coefficients while $\delta n_e$ and $\delta q_d$ denotes fluctuation electron number density and fluctuation charges of grains from their equilibrium values respectively.

## STABILITY, INSTABILITY AND ION CONVECTIONS

We assumed that ions have some convectional velocity $U_0$. In order to resolve problem of stability of DIAW, instead of to solve full fluid model of equation for dusty plasmas (continuity equations, fluid motion equations and Poisson equation ) we have to opportunity dispersion relation for ideal dusty plasmas (without fluctuations) [1]. The fluctuations of electrostatic potential $\phi$ provide to fluctuations of the number density of components of dusty plasmas. Next in results arise the fluctuations of the charges of dusty. Therefore we use 'Linear response method' [6] to solve dispersion relation for dusty plasmas with fluctuations. In order to applied mention above method we can write:

$$q_j \delta n_j (k, \omega) = -\varepsilon_0 k^2 \chi_j(k, \omega) \phi(k, \omega), \qquad (10)$$

where $\varepsilon_0$ denotes free space permittivity, $\chi_j(k, \omega)$ denotes susceptibility of j-th components of dusty plasmas without fluctuations of grain of dust, $\phi(k, \omega)$-denotes electrostatic potentials of dusty plasmas with fluctuations of grains of dust, in Fourier representation (after transformation with respect to time and length ) and $\delta n_j$ -denotes fluctuations number density of j-th components also in Fourier representations. It is relation between permittivity $\varepsilon(k,\omega)$ for dusty plasmas and susceptibility $\chi_j(k,\omega)$ of components of dusty plasmas:

$$\varepsilon(k,\omega) = \varepsilon_0\left(1 + \sum_j \chi_j(k,\omega)\right). \tag{11}$$

The permittivity $\varepsilon(k,\omega)$ designate dispersion relation for a dusty plasmas (in these paper for a plasmas without fluctuations of charges of grains ) by the equation:

$$\varepsilon(k,\omega) = 0. \tag{12}$$

Then we write Poisson equation for a dusty plasmas with fluctuations of charges of grains:

$$\varepsilon_0 k^2 \phi(k,\omega) = \sum_j q_j \delta n_j(k,\omega) + \delta q_d(k,\omega) n_{d0}, \tag{13}$$

where $n_{d0}$ denote equilibrium number density of grain of dust. Taking into account formulae (13),(10) and (11), we obtain the following equation:

$$\varepsilon(k,\omega) = \frac{\delta q_d(k,\omega)}{k^2 \phi(k,\omega)} n_{d0}. \tag{14}$$

Then, according to the formulae (5), (9), after linearization about equilibrium and Fourier transform respect time and length, we obtain:

$$\delta q_d(k,\omega) = \frac{-\nu_e \dfrac{e}{n_{d0}}\left(i\omega + \dfrac{e}{n_{d0}}\mu_e\right)\delta n_e(k,\omega)}{\left(\omega^2 + \dfrac{e^2}{n_{d0}^2}\right)}. \tag{15}$$

Now, we substitute (10), for j=e to (15) and next combined with (11) and (14). Finally, we obtain:

$$1 + \sum_j \chi_j(k,\omega) = \frac{-\nu_e \dfrac{e}{n_{d0}}\left(i\omega + \dfrac{e}{n_{d0}}\mu_e\right)\chi_e(k,\omega) n_{d0}}{e\left(\omega^2 + \dfrac{e^2}{n_{d0}^2}\mu_e^2\right)} \tag{16}$$

dispersion relation for dusty plasma with fluctuation of charges of grains, if we select proper susceptibilities $\chi_j$. Putting zero on the right hand of (16) we receive the dispersion relation for dusty plasmas without fluctuations of charges of grains. In order to receive dispersion relation for DIAW with ions convections ( with velocity $U_0$ ) and fluctuations of charges of grains we select the following susceptibilities in equation (16):

$$\chi_e(k,\omega) = \frac{1}{k^2 \lambda_{De}^2}, \quad \chi_i(k,\omega) = \frac{\omega_{pi}^2}{(\omega - kU_0)^2} \text{ (in } \chi_i \text{ Doppler shift is regard ) and } \chi_d(k,\omega) = \frac{\omega_{pd}^2}{\omega^2},$$

where $\lambda_{Dj}$ and $\omega_{pj}$ denotes respectively Debyea length and plasma frequency for j-th components of dusty plasmas. Therefore, we assume that coefficients $\nu_e$ and $\dfrac{e\mu_e}{n_{d0}}$ are small in comparison with

$\omega_{dia} = \sqrt{\dfrac{k^2 \lambda_{De}^2 \omega_{pi}^2}{1 + k^2 \lambda_{De}^2}}$ and their second order of values can be omitted. If we resolve (16) in view above remarks,

that in long wave approximation we receive:

$$\omega = kU_0 \pm k\lambda_{De}\omega_{pi} - i\frac{\lambda_{De}\omega_{pi}\nu_e}{2(\lambda_{De}\omega_{pi} \pm U_0)}. \qquad (17)$$

We have some analogy to Landau damped plasmas waves in kinetic theory: when convective and DIAW velocities have the same direction but phase velocity of wave is smaller than convective velocity we have instability. In other case the wave is damped and attenuation coefficient is proportional to $\nu_e$.

## CONCLUSIONS

Linear response method make possible us use dispersion relation for dust-ion-acoustic wave (DIAW) without fluctuation of charges of grains of dust. In the case of small fluctuations is agreement with linear approximation. Doppler shift frequency in ions susceptibility enable us use linear response method when ions have convectional velocity. We have some analogy to Landau damping: when convective velocity is larger than phase velocity of wave, if they have the same direction, we have instability. In the other case, we have linear stability of DIAW, which enable phenomenological coefficient $\nu_e$.

## ACKNOWLEDGMENTS

This research is supported by KBN grant 2PO3B-126- 24